\documentclass[prl,
                preprint,
                amsmath,
                amssymb,
                showpacs,
                superscriptaddress]{revtex4}
\usepackage{graphicx} 
\usepackage{dcolumn}  
\usepackage{bm}       
\usepackage{amsfonts}
\usepackage{dsfont}
\usepackage{soul,xcolor}
\usepackage{dirtytalk}
\usepackage{ulem}
\usepackage{color}
 \newcommand{\Nadd}[1]{\textcolor{red}{#1}}

\begin{document}


\title{Signatures of the orbital angular momentum of an infrared light beam in the two-photon transition matrix element: A step toward attosecond chronoscopy of photoionization}
 
\author{Sucharita Giri}
\affiliation{%
Department of Physics, Indian Institute of Technology Bombay,
            Powai, Mumbai 400076  India}

\author{Misha Ivanov}
\affiliation{%
Max-Born-Institut, Max Born Strasse 2A, 12489 Berlin, Germany }
\affiliation{%
Blackett Laboratory, Imperial College London, South Kensington Campus, SW7 2AZ London, United Kingdom}
\affiliation{%
Department of Physics, Humboldt University, Newtonstrasse 15, 12489 Berlin, Germany}

\author{Gopal Dixit}
\email[]{gdixit@phy.iitb.ac.in}
\affiliation{%
Department of Physics, Indian Institute of Technology Bombay,
            Powai, Mumbai 400076  India}




\begin{abstract}
We present a theory of time-resolved photoionisation in the presence of
a vortex beam. In a  pump-probe setup, an extreme ultraviolet or an x-ray pump pulse 
triggers ionization, which is probed by a synchronized infrared pulse with non-zero
orbital angular momentum. We show, how this property of the probe pulse affects
the electron dynamics upon ionization, in a way that is independent of 
the initial and final angular momentum states of the ionizing system. 
\end{abstract}

\maketitle
\section{Introduction}

In their pioneering work, Allen and co-workers have  shown the existence of the  
orbital angular momentum (OAM) of a light beam, along with its spin angular  momentum~\cite{allen1992orbital}.  
The OAM of light is related to the spatial profile of the light beam and 
is characterized by the topological charge $l$, 
which can have only integer values from 0 to $\pm \infty$,   
whereas the spin angular momentum  is related to the polarization properties of light. 
Since its first demonstration,  the OAM of light has found numerous macroscopic 
applications in 
different fields ~\cite{cardano2015spin, torres2011twisted, babiker2018atoms}
including quantum information processing~\cite{wang2012terabit}, 
optical interferometry~\cite{furhapter2005spiral}, 
chiral recognition in molecules~\cite{forbes2019raman, brullot2016resolving, forbes2018optical},  
manipulation of nanoparticles~\cite{yao2011orbital} and in quantum gases~\cite{andersen2006quantized, inoue2006entanglement}. 
While it is relatively straightforward to generate beams with non-zero orbital angular momentum 
in the infrared (IR) and visible range, similar technology is much more difficult in the extreme ultraviolet (XUV) and 
x-ray range. This has stimulated recent works on using high-harmonic generation 
to up-convert OAM beams from the IR to the XUV and soft x-ray energy range
~\cite{zurch2012strong, hernandez2013attosecond, gariepy2014creating, geneaux2016synthesis, rego2016nonperturbative, turpin2017extreme, hernandez2017extreme, paufler2018tailored, gauthier2019orbital}.   
An elegant non-collinear scheme has been employed to generate linearly polarized twisted 
light beams with relatively low OAM~\cite{gauthier2017tunable, kong2017controlling, dorney2019controlling}
via high-harmonic generation. 
These breakthroughs  have opened a door to synthesize 
coherent attosecond XUV and x-ray pulses with  desirable OAM properties, adding  
an additional knob to the attosecond light generation toolbox. 
 
It may appear that, apart from the space-dependent amplitude and phase modulation, 
the macroscopic OAM properties of an IR or visible light beam will hardly manifest  
at the single-atom, microscopic level due to the difference in the spatial scales.
As long as the dipole approximation dominates the atomic response, each atom sees a 
local field at a given spatial position inside the OAM beam. However,
this general impression is not always accurate, as shown in particular
 in \cite{schmiegelow2016transfer}, using a Bose-Einstein condensate
with naturally delocalized centre-of-mass atomic wavefunction.  
Here we analyse an alternative possibility and show how the 
topological 
charge of an incident OAM IR beam can be encoded via attosecond chronoscopy of 
photoionisation~\cite{pazourek2015attosecond} by a pair of time-synchronized
x-ray pump and IR probe pulses.  
Attosecond streaking and RABBITT (reconstruction of attosecond beating by interference of two-photon transitions) are the most common experimental methods to
perform attosecond chronoscopy of  photoionization~\cite{pazourek2015attosecond}. 

\begin{figure}[t!]
\includegraphics[width=10 cm]{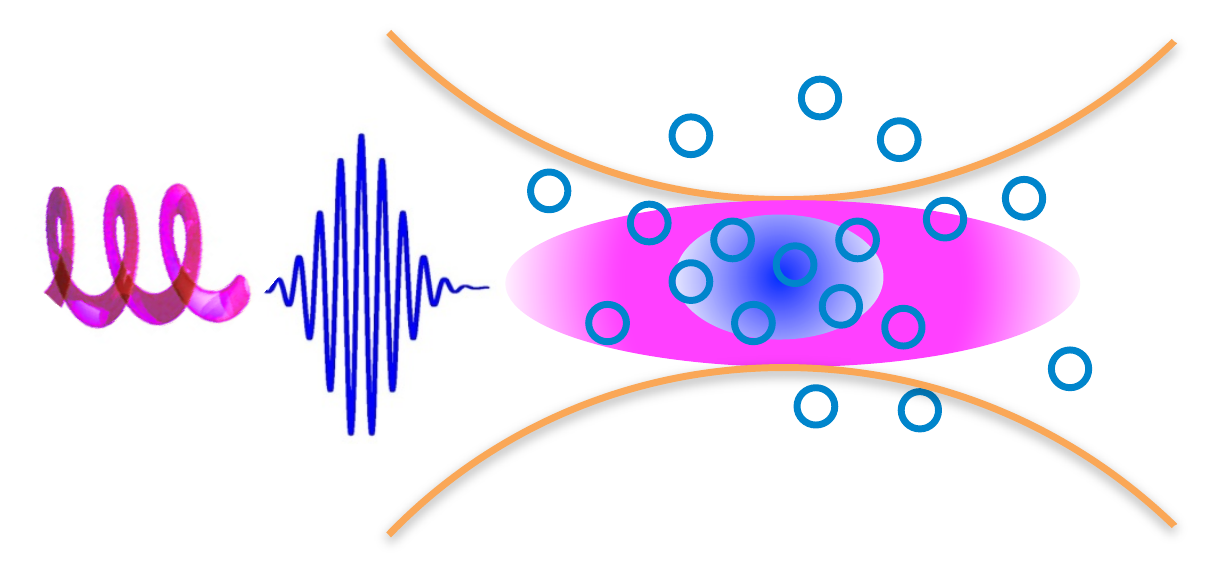}
\caption{Schematic of time-resolved photoionization by a pair of regular x-ray 
pump pulse with zero orbital angular momentum  (shown in blue)  and a twisted
infrared probe pulse  (shown in pink). The focal volumes of both pulses are shown by their 
respective colours, atoms are depicted with blue open circles. The x-ray pump pulse ionizes 
the electrons from the atoms and the liberated  photoelectrons, within the focal volume,   
are probed by the twisted infrared pulse.} 
\label{fig1}
\end{figure}

In our proposed scheme, an attosecond  x-ray pump pulse
ionises atoms (shown in blue colour in Fig.~\ref{fig1}). 
The liberated photoelectrons are probed by a twisted IR probe pulse (shown in pink). 
The short wavelength of the pump pulse allows one to focus it 
tightly into the centre of the OAM doughnut, 
where the amplitude and the phase structure of the twisted IR probe
can be experienced by the photoelectrons. 
In the generalised theory of time-resolved photoionisation,  
both the pump and probe pulses can carry OAM. However, for our purpose, we limit the 
analysis to the case when the X-ray pump pulse has no OAM but the IR probe can carry OAM. 
The fact that the X-ray pump pulse does not have to carry OAM, 
simplifies practical realisation of
the proposed setup using currently developing X-ray sources.

As shown by Picon and co-workers ~\cite{picon2010transferring, picon2010photoionization}, 
when the phase and amplitude inhomogeneity of a twisted beam becomes important, the standard dipolar 
selection rules are no longer adequate~\cite{babiker2002orbital}. 
 Exchange of angular momentum larger than 
one unit has been found during photoionisation~\cite{picon2010transferring, picon2010photoionization}
and photoexcitation of valence electrons with twisted beams~\cite{peshkov2017photoexcitation, schmiegelow2016transfer, giammanco2017influence, babiker2018atoms}. 
Attosecond time-delay in photoionisation has been predicted 
to be sensitive to the magnetic sublevels of a spherically symmetric atom 
ionised by a twisted XUV beam~\cite{watzel2016discerning}. 

There is an important difference between 
all-optical setups, which detect light generated by matter interacting
with OAM beams, and setups detecting photoelectrons.
In the former case,  the OAM properties of the incident light are recorded across a macroscopic 
sample and affect the measurement via coherent addition of the optical response
from different parts of the interaction region. 
While measuring photoelectrons,  OAM-sensitive signal
arises when an atom is placed near the centre of
the twisted beam, where  amplitude and phase  modulation is maximised (at the expense of 
low incident field intensity).

\section{Theory}
In the  time-resolved photoionization, within the pump-probe scenario,  the 
pump pulse liberates an electron from the ground state $| i \rangle$ 
with negative energy $\epsilon_{i}$. The electron is promoted  
to a  continuum state $| k^{\prime} \rangle$ 
with energy $\epsilon_{k^{\prime}}$. The photoelectron is probed by the
twisted IR pulse, which induces a transition from a  continuum state $| k^{\prime} \rangle$ 
to another continuum state $| k \rangle$ with energy $\epsilon_{k}$. 
In the following, we present the theory of time-resolved photoionization. 

First, we write the two-photon transition amplitude 
corresponding to the absorption of both one pump and one probe photons, 
\begin{equation}\label{eq01}
M_{\epsilon_{k},\epsilon_{i}}^{(2)} =  i~\lim_{\delta \to 0}\int_{0}^{\infty} d\epsilon_{k^{\prime}} 
\frac{ \langle k | \mathcal{H}_{\textrm{int}} | k^{\prime} \rangle 
\langle k^{\prime} | \mathcal{H}_{\textrm{int}} | i \rangle }
{(\epsilon_{k^{\prime}}-\epsilon_{i}-\Omega-i\delta)}. 
\end{equation}
Here, the energy of the final continuum state
is $\epsilon_{k} = \epsilon_{i} + \Omega + \omega$, where
 $ \Omega$ and $\omega$ are the photon energies of the pump and IR probe pulses, respectively. 
The pump photon energy  is larger than the ionization potential of  the atomic system.
  
The interaction Hamiltonian  is 
$\mathcal{H}_{\textrm{int}} = -\mathbf{\hat{A}}(\mathbf{r}, t)\cdot \mathbf{\hat{p}}$ 
where $\mathbf{\hat{A}}(\mathbf{r}, t)$ is the vector potential of the pump and IR probe pulses.
The vector potential corresponding to the probe vortex pulse, which induces 
a transition from $| k^{\prime} \rangle$ to $| k \rangle$, 
has the following form~\cite{picon2010photoionization} 
\begin{equation}
\label{eq02}
\mathbf{A} (\mathbf{r},t) = \mathbf{A}_{0} w_{0}~g(t)~\left[e^{i (\mathbf{k} \cdot \mathbf{r}-  \omega t)} ~\textrm{LG}_{l,p} (\rho, \phi, z; \mathbf{k}) + \textrm{c.c.} \right].
\end{equation} 
Here, $\mathbf{A}_{0}$ is the amplitude (including polarization state), $w_{0}$ is the beam waist of the pulse, 
$g(t)$ is the envelope of the pulse,  $\mathbf{k}$ is a carrier wave vector, 
$\omega$ is carrier frequency and  $\textrm{LG}_{l,p} $ is the Laguerre-Gaussian modes of the pulse, 
which contains the transverse spatial structure of the vortex pulse with $l$ and $p$ corresponding 
to the topological charge and the radial node of the LG mode, respectively.  
In the present work, we have used the expression of $\textrm{LG}_{l,p} (\rho, \phi, z; \mathbf{k})$  as given in Eq. (2) of Ref.~\cite{picon2010photoionization}. 
 
The ground  state wavefunction is written as a 
product of the radial and angular parts, 
$\varphi_{i}(\mathbf{r}) = \langle \mathbf{r} | i \rangle = 
R_{n_{i},l_{i}}(r) Y_{l_{i},m_{i}}(\hat{r})$. The  final 
continuum state wavefunction is expressed in terms of the
partial wave expansion as 
$\varphi_{k}(\mathbf{r}) = (8\pi)^{\frac{3}{2}}\sum_{l,m} ~i^{l}~e^{-i\eta_{l}} 
R_{k, l}(r) Y_{l,m}(\hat{r})Y_{l, m}^{*}(\hat{k})$. 
The phase  $\eta_{l}$ is the scattering phase, i.e., the difference between
the phase of the photoelectron ejected from the atom and that of a free-electron.
 
We substitute these expressions in the total two-photon amplitude Eq.~({\ref{eq01}}), 
which reduces  to the product of radial and angular parts
\begin{equation}
\label{eq03}
M_{\epsilon_{k},\epsilon_{i}}^{(2)}  \propto  i(8\pi)^{\frac{9}{2}} A_{\omega} A_{\Omega}
\sum_{l, m} \sum_{l^{\prime}, m^{\prime}} (-i)^{l} e^{i\eta_{l}} Y_{l, m}( \hat{k} )~
T_{l, l^{\prime}, l_{i}}  \times \textrm{Angular part},
\end{equation}
where $A_{\Omega}$ and $A_{\omega}$ are, respectively, the amplitudes of the vector potentials of 
the pump and probe pulses and  $T_{l, l^{\prime}, l_{i}}$ is the radial part of the 
two-photon amplitude. 
The quantum numbers $l_{i}$, $l^{\prime}$ and $l$ are the angular momenta of the initial ground, 
intermediate continuum and final continuum states. 
The range of accessible  angular momenta for intermediate and final continuum states
is decided by the selection rules. For vortex pulses, the selection rules are 
significantly different compared to the standard dipole selection rules 
~\cite{picon2010transferring, picon2010photoionization, babiker2002orbital, babiker2018atoms}. 
Here we assume that both pulses are linearly polarised along 
the quantization axis $z$.
 
The radial part of the two-photon transition amplitude 
$M_{\epsilon_{k},\epsilon_{i}}^{(2)}$ is
\begin{equation}\label{eq04}
T_{l, l^{\prime}, l_{i}}  = \lim_{\delta \to 0} \int_{0}^{\infty}
d\epsilon_{k^{\prime}} \frac{\langle R_{k, l} | \mathcal{H}_{\textrm{int}}(r) | R_{k^{\prime}, l^{\prime}} \rangle 
\langle R_{k^{\prime}, l^{\prime}} | \mathcal{H}_{\textrm{int}}(r) | R_{{n_{i}, l_{i}}} \rangle}{(\epsilon_{i}+\Omega - \epsilon_{k^{\prime}}+ i \delta)},
\end{equation}
where $\mathcal{H}_{\textrm{int}}(r)$ is the radial part of the interaction Hamiltonian. 
We first compute the continuum-continuum (CC) transition
amplitude, $\langle R_{k, l} | \mathcal{H}_{\textrm{int}}(r) | R_{k^{\prime}, l^{\prime}} \rangle$,  
and then substitute the result for  $\langle R_{k, l} | \mathcal{H}_{\textrm{int}}
(r) | R_{k^{\prime}, l^{\prime}} \rangle$ into the energy integral of Eq.~({\ref{eq04}}),  to 
obtain the compact form of $T_{l, l^{\prime}, l_{i}}$. 
We use the standard asymptotic form of the radial continuum wavefunction,  
\begin{equation}\label{eq05}
\lim_{r \to \infty} R_{k ,l}(r) = \frac{1}{r} \sqrt{\frac{2}{\pi k}}~\textrm{sin} \left[kr - \frac{\pi l}{2} +\eta_{l}(k) +Z \frac{\textrm{ln} (2kr)}{k} \right], 
\end{equation} 
where $\eta_{l}(k)$ is the scattering phase-shift. 
It can be written as $ \eta_{l}(k) = \sigma_{l}(k) + \delta_{l}(k) $ 
with $\sigma_{l}(k) = \text{arg}[\Gamma(l+1-i Z/k)]${\color{red},} the Coulomb phase-shift 
and $\delta_{l}(k)$ originating from the short-range deviations of the ionic potential 
from a purely 
Coulombic form. 
In the asymptotic region, the phase of the continuum wavefunction also 
exhibits the standard logarithmic divergence characteristic for the Coulomb potential of the 
ionic core with charge $Z$. Note that the chosen form of the continuum wavefunction 
is not a good choice at low energy as shown in Ref.~\cite{dahlstrom2013theory}.

By using the expression of radial continuum wavefunction from Eq.~(\ref{eq05}), 
the CC radial matrix element is 
\begin{eqnarray}\label{eqextra}
\langle R_{k, l} | \mathcal{H}_{\textrm{int}}(r) | R_{k^{\prime}, l^{\prime}} \rangle  & = &
\frac{2}
{\pi \sqrt{k k^{\prime}}} \left( \frac{\sqrt{2} }{w_{0}} \right)^{\vert l_{2} \vert} \int dr  r^{2} \frac{1}{r^{2}} \textrm{sin}\left[kr - \frac{\pi l}{2} +\eta_{l} (\epsilon_{k}) +Z \frac{ \textrm{ln} (2kr)}{k}\right] 
\nonumber \\ 
&& \times  (r)^{1+ \vert l_{2} \vert}  \textrm{sin} \left[k^{\prime}r - \frac{\pi l^{\prime}}{2} +\eta_{l^{\prime}} (\epsilon_{k^{\prime}}) +Z \frac{\textrm{ln}  (2k^{\prime} r)}{k^{\prime}} \right].   
\end{eqnarray}
Here,  $\mathcal{H}_{\textrm{int}}(r) $ is responsible 
for the CC transitions and depends on $|l_{2}|$, which is the topological charge of the vortex IR pulse. 
In the calculations,  we have only retained the dominant term proportional 
to $\cos (k-k^{\prime})$  and ignored the term proportional to $\cos (k+k^{\prime})$.
The simplified expression for the CC radial matrix element is 
\begin{eqnarray}\label{eq06}
\langle R_{k, l} | \mathcal{H}_{\textrm{int}}(r) | R_{k^{\prime}, l^{\prime}} \rangle  & = & \frac{1}
{\pi \sqrt{k k^{\prime}}} \left( \frac{\sqrt{2} }{w_{0}} \right)^{\vert l_{2} \vert}
\lim_{\zeta \to 0} e^{i \Delta^{\prime}} 
\frac{(2k^{\prime})^{\frac{iZ}{k^{\prime}}}}{(2k)^{\frac{iZ}{k}}}  
\left( \frac{i}{k'-k-i \zeta} \right)^{2+ \vert l_{2} \vert +iZ \left(\frac{1}{k'} - \frac{1}{k} \right) } 
\nonumber \\ 
&& \times~ \Gamma \left[ 2+ \vert l_{2} \vert +iZ \left(\frac{1}{k'} - \frac{1}{k} \right)  \right] +\textrm{c. c.}
\end{eqnarray}
Here, 
$ \Delta= \eta_{l^{\prime}} (k^{\prime}) -\eta_{l} (k)$ 
and $ \Delta^{\prime} = \Delta + \frac{\pi}{2} (l-l^{\prime})$. 

Substituting  Eq.~({\ref{eq06}}) in Eq.~({\ref{eq04}}) and performing 
the complex energy integral near the pole corresponding
to the energy $\epsilon_{\kappa} = \kappa^{2}/2 = \epsilon_i+\Omega$ yields 
the compact expression for $T_{l, l^{\prime}, l_{i}}$ 
\begin{eqnarray}\label{eq07}
T_{l, l^{\prime}, l_{i}}  & = & \frac{-i}{\sqrt{k \kappa}} 
\left( \frac{\sqrt{2} }{w_{0}} \right)^{\vert l_{2} \vert} 
\langle R_{\kappa, l^{\prime}} | \mathcal{H}_{\textrm{int}}(r) | R_{{n_{i}, l_{i}}} \rangle
\left( \frac{1}{\vert \kappa - k \vert}\right)^{2+ \vert l_{2} \vert} \exp \left[- \frac{\pi Z}{2} \left( \frac{1}{\kappa} - \frac{1}{k} \right) \right]
 \nonumber \\
&& \times (i)^{\vert l_{2} \vert }   e^{i \Delta^{\prime}} \frac{(2\kappa)^{\frac{iZ}{\kappa}}}{(2k)^{\frac{iZ}{k}}}\left( \frac{1}{\kappa - k } \right)^{iZ\left( \frac{1}{\kappa} - \frac{1}{k} \right)}  
\Gamma \left[ 2+ \vert l_{2} \vert +iZ \left(\frac{1}{\kappa} - \frac{1}{k} \right) \right].
\end{eqnarray}
In the above equation, the amplitude and phase terms are separated. 
The first line contains bound-continuum one-photon transition amplitude, whereas 
the second line \Nadd{contains} the phase term of the CC transition amplitude whose strength 
is controlled by the exponential term. 
  
After substituting the result of Eq.~({\ref{eq07}}) in  Eq.~({\ref{eq03}}), the 
total two-photon transition amplitude reduces as
\begin{eqnarray}\label{eq08}
M^{(2)}_{\epsilon_{k},\epsilon_{i}} & \propto & -  \frac{1}{\sqrt{k \kappa}} (8\pi)^{\frac{9}{2}} A_{\omega} A_{\Omega}   \left( \frac{\sqrt{2} }{w_{0}} \right)^{\vert l_{2} \vert} \exp \left[- \frac{\pi Z}{2} \left( \frac{1}{\kappa} - \frac{1}{k} \right) \right] \nonumber \\
&& \times \frac{1}{\vert k - \kappa \vert^{2+ \vert l_{2} \vert}}  \ \frac{(2\kappa)^{\frac{iZ}{\kappa}}}{(2k)^{\frac{iZ}{k}}}  
\left( \frac{1}{\kappa-k} \right)^{ iZ \left(\frac{1}{\kappa} - \frac{1}{k} \right) } 
\ \Gamma \left[ 2+ \vert l_{2} \vert +iZ \left(\frac{1}{\kappa} - \frac{1}{k} \right) \right]  \nonumber \\
&& \times \sum_{l, m} \sum_{l^{\prime}, m^{\prime}} \ (i)^{-l^{\prime} + \vert l_{2} \vert} e^{i\eta_{l^{\prime}}(\kappa)}\  Y_{l, m}( \hat{k} )\ 
\langle R_{\kappa, l^{\prime}} | \mathcal{H}_{\textrm{int}}(r) | R_{{n_{i}, l_{i}}} \rangle .
\end{eqnarray}
The phase of the total amplitude in the asymptotic limit is
\begin{equation}\label{eq09}
\textrm{arg} \left[ M^{(2)}_{\epsilon_{k},\epsilon_{i}}\right]  =  
\pi + \textrm{arg}\left[ Y_{l, m}(\hat{k}) \right] + \phi_{\Omega} + \phi_{\omega} 
 - \frac{\pi}{2}(l^{\prime} - \vert l_{2} \vert ) + \eta_{l^{\prime}(\kappa)} + \phi_{cc}(k, \kappa),
\end{equation}
where,  $\phi_{\Omega}$ and $\phi_{\omega}$ are the phases of the pump
and probe pulses; respectively.  
The quantum phase associated with the CC transition is
\begin{equation}\label{eq10}
\phi_{cc} \left(k , \kappa \right) = \textrm{arg}\left[ \frac{(2\kappa)^{\frac{iZ}{\kappa}}}{(2k)^{\frac{iZ}{k}}}  \times \frac{\Gamma \left[ 2 + \vert l_{2} \vert +iZ (\frac{1}{\kappa} - \frac{1}{k})\right] }{\left(\kappa -k \right)^{iZ(\frac{1}{\kappa} - \frac{1}{k})}}\right]. 
\end{equation}
Note that, $\phi_{cc}$ originates from the absorption of  the vortex  IR pulse 
in the presence of the Coulomb potential $Z$. 
It is important to notice that the 
vortex nature of the IR pulse $(\vert l_{2} \vert)$ 
appears in the argument of the Gamma function, independently of the  
initial atomic state.  If one substitutes $l_{2} = 0$, the expression of $\phi_{cc}$ 
reduces to the known expression as reported in Eq. (22) in Ref.~\cite{dahlstrom2013theory}. 

\begin{figure}[t!]
\includegraphics[width=12 cm]{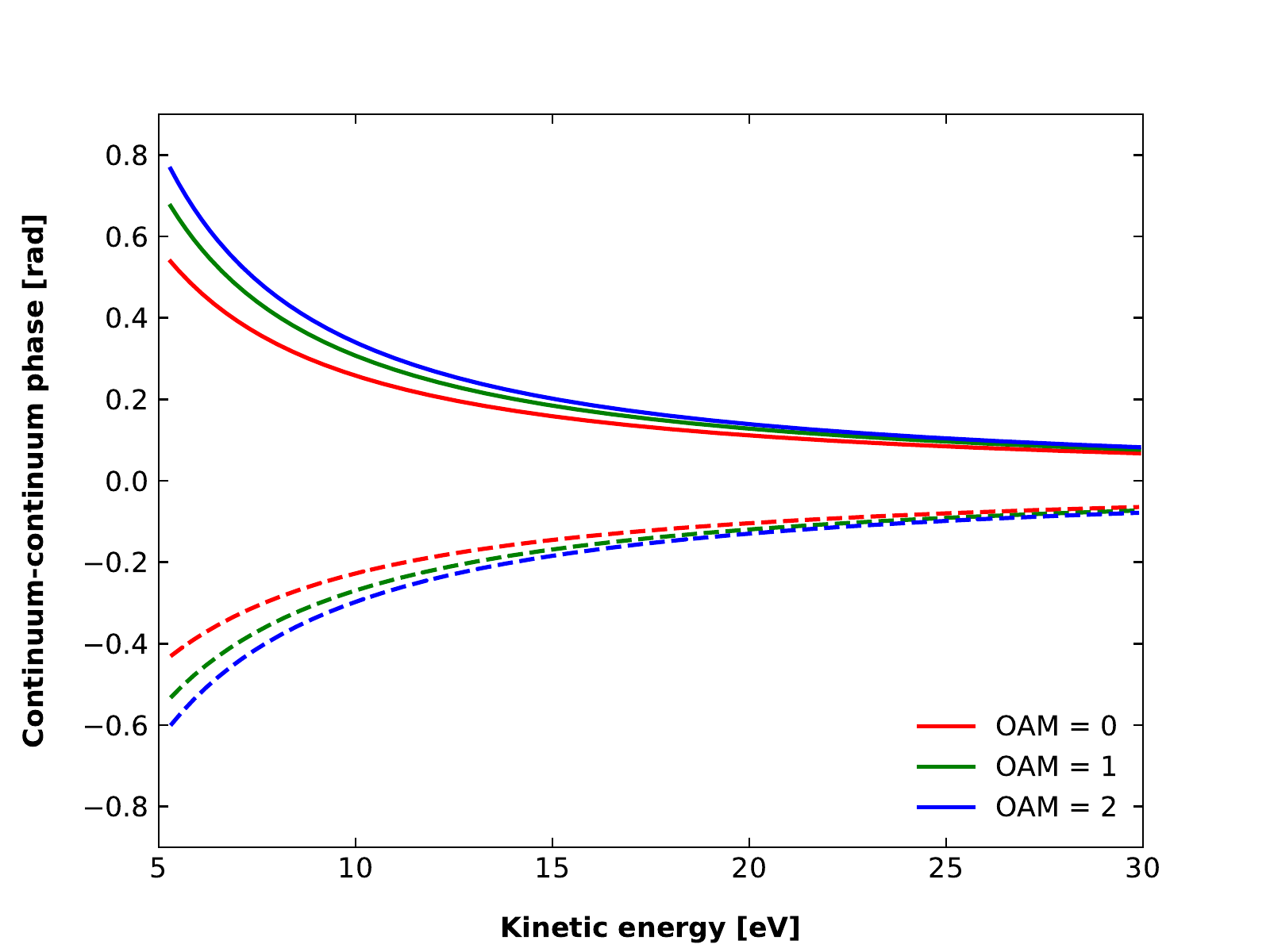}
\caption{The continuum-continuum quantum phases corresponding to 
absorption (solid lines) and emission (dotted lines) of the vortex IR pulse with
 different OAM values (topological charge $|l_{2}|$) for 
 $\omega = 1.55$ eV and $Z =1$. Here, OAM = 0 corresponds to regular IR pulse. The legend represents the solid curves from bottom to top and the dotted curves from top to bottom.} 
\label{phase}
\end{figure} 

The calculation of the two-photon transition matrix element corresponding to the emission of
the IR photon is similar,  only the phase of the CC transition is opposite
to the case of the IR absorption.

The CC-phases for the different values of the OAM of the IR pulse are presented in Fig.~\ref{phase}. 
The solid and dotted lines correspond to the absorption and emission of an IR photon, 
for the same final kinetic energy, respectively.   
The phases  
are positive for absorption and negative for emission;  whereas 
the magnitudes of the CC-phases are not exactly equal 
as evident from the figure. 
For low kinetic energies of the photoelectron, the phases increase with increasing the OAM.  
For high energies, the phases become insensitive to the OAM 
and approach those for regular light with zero OAM, as reflected in the figure.
 
The insensitivity of $\phi_{cc}$ to OAM for high kinetic energies follows from the fact that the dominant complex-valued and energy-dependent
part of the two-photon transition amplitude comes from the phase of the Gamma-function. 
Since  $(\frac{1}{\kappa} - \frac{1}{k}) \rightarrow 0$ as $k 
\rightarrow \infty$,  $iZ(\frac{1}{\kappa} - \frac{1}{k}) 
\rightarrow 0$ in the argument of the Gamma function, which becomes real-valued.

\section{Results and Discussion}

An experimental way to probe the quantum phase associated 
with CC-transition is offered by the 
RABBITT method. In this method, an XUV or x-ray 
attosecond pulse train is used as a pump pulse, with the photoelectron  
probed by the IR pulse.  
The pump-probe delay-dependent changes in the photoelectron spectrum give access to 
the sub-IR-cycle temporal resolution. In RABBITT, the same final state can be reached by two paths: either 
an absorption of pump photon 
[($2q-1$)-th harmonic] followed by an absorption of an IR photon 
($\omega \tau = \phi_{\omega}$) or an absorption of pump photon 
[($2q+1$)-th harmonic] followed by an emission of 
an IR photon with phase ($-\omega \tau = -\phi_{\omega}$). 
Both quantum paths lead to the same final `sideband' state at the energy of 
the $2q$-th harmonics. 
The phase-dependent  interferences of the two paths yield the oscillations in the 
intensity of the sideband, which is  
written in terms of the interference of the two-photon transition amplitudes corresponding 
to absorption $M^{(a)}$ and emission of an IR photon $M^{(e)}$ as $\mathcal{I}_{2q} = |M^{(a)} + M^{(e)}|^{2}$. The interference-term of $\mathcal{I}_{2q}$ encodes the phase as
\begin{equation}\label{eq11}
\text{arg}(M^{(a)} M^{(e)})  \approx  -2\omega \tau + \phi_{2q+1} -\phi_{2q-1} 
+ \eta_{l^{\prime}}(\kappa_{<}) - \eta_{l^{\prime}}(\kappa_{>}) + \phi_{cc}(k,\kappa_{>}) - \phi_{cc}(k,\kappa_{<}). 
 \end{equation}
Here ($-2\omega \tau + \phi_{2q+1} -\phi_{2q-1}$)-term contains the phase of the field. The 
finite-difference version of the Wigner-Smith time-delay is expressed in terms of the scattering phase: 
$\tau_{l^{\prime}}^{\textrm{WS}}(k) = [\eta_{l^{\prime}}(\kappa_{<}) - \eta_{l^{\prime}}(\kappa_{>})]/(2\omega)$
and the time-delay corresponding to the CC-phase is 
 \begin{equation}\label{eq13}
\tau_{cc}(k) \approx \frac{\phi_{cc}(k,\kappa_{>}) - \phi_{cc}(k,\kappa_{<})}{2\omega}.
\end{equation} 
The CC time delay is the finite difference approximation for the derivative of $\phi_{cc}$ with respect to 
2$\omega$ in RABBITT. 
It can be seen that $\tau_{cc}$ is universal 
and  independent of the initial and final angular momenta states of the atom. 
Here, $\epsilon_{<} = \epsilon_{i}+(2q-1)\omega = \kappa_{<}^{2}/2$ and $\epsilon_{>} = \epsilon_{i}+(2q+1)\omega = 
\kappa_{>}^{2}/2$ are the energies of the two quantum paths. 

\begin{figure}[t!]
\includegraphics[width=12 cm]{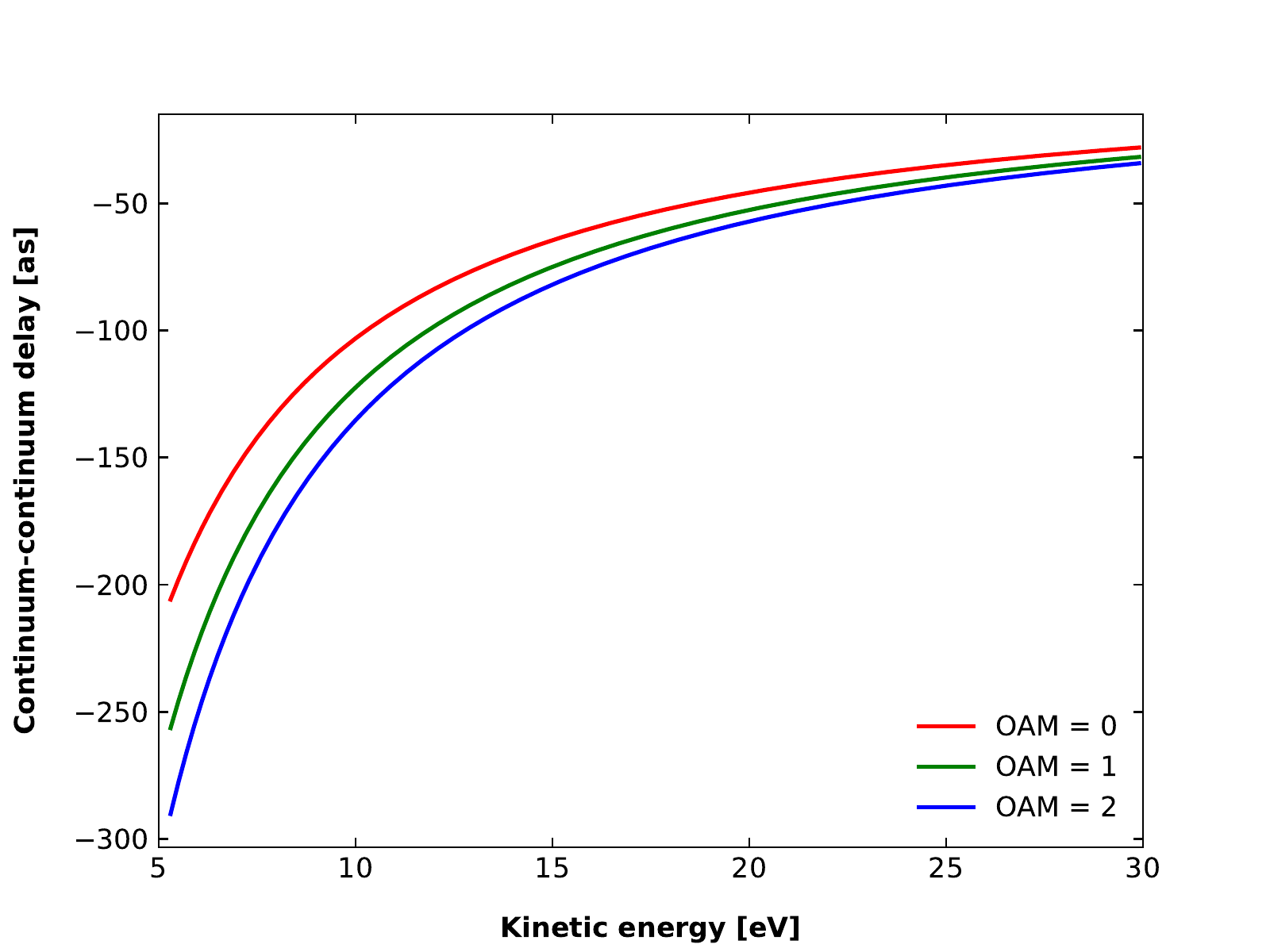}
\caption{
The time-delays corresponding to continuum-continuum  transitions, $\tau_{cc}$, 
for different OAM values of the vortex IR pulse with $\omega = 1.55$ eV and $Z = 1$.
Here, OAM = 0 corresponds to regular IR pulse. The legend represents the solid curves from bottom to top and the dotted curves from top to bottom.} 
\label{delay}
\end{figure} 

Figure~\ref{delay} presents $\tau_{cc}$  for different values of OAM of vortex IR pulse. 
In general, $\tau_{cc}$ increases monotonically as a function of kinetic energy but remains 
negative: the photoelectron appears to be ahead compared to the electron moving in
the absence of the core potential. 
In the relatively low kinetic energy regime, $\tau_{cc}$ becomes more negative for a 
larger value of OAM. 
The delays merge to a constant value as the kinetic energy increases. 

\section{Conclusion}
In conclusion, we have  
developed the theory of time-resolved photoionization induced by a vortex pulse carrying OAM. The main finding 
of this study is the impact of the OAM of the probe beam on  photoionization delays. As in a standard RABBITT setup, the total 
phase associated with the ionization process is a sum of the two phases, the scattering phase associated with the bound-free 
transition induced by the pump  pulse and the 
 CC phase associated with the transition induced by the IR pulse.  
The CC phase depends parametrically on the OAM of the vortex pulse and 
 the charge of the remaining ionic core. 
The CC phase 
does not depend on the angular momentum states of the system and the short-range 
behaviour of the atomic potential. 
Our theory can also be used for other interferometric photoionization 
measurement methods such as  the 
attosecond streaking camera for relatively low IR intensity as discussed in Ref.~\cite{dahlstrom2013theory}. The present work provides a  step 
towards attosecond chronoscopy of photoionisation in twisted light beams. 
Other
important effects such as spatial sampling for specific focusing conditions, 
better description of the
continuum wavefunction to include electron-electron correlations (e.g., in the vicinity of autoionizing states) etc. 
need to be incorporated  in specific experiments for simulations of RABBITT and streaking spectra.

\section{Acknowledgements}
G.D. acknowledges support from Science and Engineering Research Board (SERB) India 
(Project No. ECR/2017/001460) and Max-Planck India visiting fellowship. 
M.I. acknowledges DFG QUTIF program, Grant IV 152/6-2.


\begin{thebibliography}{10}

\bibitem{allen1992orbital}
L.~Allen, M.~W. Beijersbergen, R.~J.~C. Spreeuw, and J.~P. Woerdman,
\newblock Phys. Rev. A {\bf 45}, 8185 (1992).

\bibitem{cardano2015spin}
F.~Cardano and L.~Marrucci,
\newblock Nature Photonics {\bf 9}, 776 (2015).

\bibitem{torres2011twisted}
J.~P. Torres and L.~Torner,
\newblock {\em Twisted photons: applications of light with orbital angular
  momentum},
\newblock John Wiley \& Sons, 2011.

\bibitem{babiker2018atoms}
M.~Babiker, D.~L. Andrews, and V.~E. Lembessis,
\newblock Journal of Optics {\bf 21}, 013001 (2018).

\bibitem{wang2012terabit}
J.~Wang et~al.,
\newblock Nature Photonics {\bf 6}, 488 (2012).

\bibitem{furhapter2005spiral}
S.~F{\"u}rhapter, A.~Jesacher, S.~Bernet, and M.~Ritsch-Marte,
\newblock Optics Letters {\bf 30}, 1953 (2005).

\bibitem{forbes2019raman}
K.~A. Forbes,
\newblock Phys. Rev. Letts. {\bf 122}, 103201 (2019).

\bibitem{brullot2016resolving}
W.~Brullot, M.~K. Vanbel, T.~Swusten, and T.~Verbiest,
\newblock Science advances {\bf 2}, e1501349 (2016).

\bibitem{forbes2018optical}
K.~A. Forbes and D.~L. Andrews,
\newblock Optics Letters {\bf 43}, 435 (2018).

\bibitem{yao2011orbital}
A.~M. Yao and M.~J. Padgett,
\newblock Advances in Optics and Photonics {\bf 3}, 161 (2011).

\bibitem{andersen2006quantized}
M.~F. Andersen, C.~Ryu, P.~Clad{\'e}, V.~Natarajan, A.~Vaziri, K.~Helmerson,
  and W.~D. Phillips,
\newblock Phys. Rev. Letts. {\bf 97}, 170406 (2006).

\bibitem{inoue2006entanglement}
R.~Inoue, N.~Kanai, T.~Yonehara, Y.~Miyamoto, M.~Koashi, and M.~Kozuma,
\newblock Physical Review A {\bf 74}, 053809 (2006).

\bibitem{zurch2012strong}
M.~Z{\"u}rch, C.~Kern, P.~Hansinger, A.~Dreischuh, and C.~Spielmann,
\newblock Nature Physics {\bf 8}, 743 (2012).

\bibitem{hernandez2013attosecond}
C.~Hern{\'a}ndez-Garc{\'\i}a, A.~Pic{\'o}n, J.~San~Rom{\'a}n, and L.~Plaja,
\newblock Phys. Rev. Letts. {\bf 111}, 083602 (2013).

\bibitem{gariepy2014creating}
G.~Gariepy, J.~Leach, K.~T. Kim, T.~J. Hammond, E.~Frumker, R.~W. Boyd, and
  P.~B. Corkum,
\newblock Phys. Rev. Lett. {\bf 113}, 153901 (2014).

\bibitem{geneaux2016synthesis}
R.~G{\'e}neaux, A.~Camper, T.~Auguste, O.~Gobert, J.~Caillat, R.~Ta{\"\i}eb,
  and T.~Ruchon,
\newblock Nature Communications {\bf 7}, 12583 (2016).

\bibitem{rego2016nonperturbative}
L.~Rego, J.~San~Rom{\'a}n, A.~Pic{\'o}n, L.~Plaja, and
  C.~Hern{\'a}ndez-Garc{\'\i}a,
\newblock Phys. Rev. Letts. {\bf 117}, 163202 (2016).

\bibitem{turpin2017extreme}
A.~Turpin, L.~Rego, A.~Pic{\'o}n, J.~San~Rom{\'a}n, and
  C.~Hern{\'a}ndez-Garc{\'\i}a,
\newblock Scientific Reports {\bf 7}, 43888 (2017).

\bibitem{hernandez2017extreme}
C.~Hern{\'a}ndez-Garc{\'\i}a, A.~Turpin, J.~San~Rom{\'a}n, A.~Pic{\'o}n,
  R.~Drevinskas, A.~Cerkauskaite, P.~G. Kazansky, C.~G. Durfee, and {\'I}.~J.
  Sola,
\newblock Optica {\bf 4}, 520 (2017).

\bibitem{paufler2018tailored}
W.~Paufler, B.~B{\"o}ning, and S.~Fritzsche,
\newblock Physical Review A {\bf 98}, 011401 (2018).

\bibitem{gauthier2019orbital}
D.~Gauthier et~al.,
\newblock Optics Letters {\bf 44}, 546 (2019).

\bibitem{gauthier2017tunable}
D.~Gauthier, P.~R. Ribi{\v{c}}, G.~Adhikary, A.~Camper, C.~Chappuis, R.~Cucini,
  L.~F. DiMauro, G.~Dovillaire, F.~Frassetto, R.~G{\'e}neaux, P.~Miotti,
  L.~Poletto, B.~Ressel, C.~Spezzani, M.~Stupar, T.~Ruchon, and G.~D. Ninno,
\newblock Nature Communications {\bf 8}, 14971 (2017).

\bibitem{kong2017controlling}
F.~Kong, C.~Zhang, F.~Bouchard, Z.~Li, G.~G. Brown, D.~H. Ko, T.~J. Hammond,
  L.~Arissian, R.~W. Boyd, E.~Karimi, and P.~B. Corkum,
\newblock Nature Communications {\bf 8}, 14970 (2017).

\bibitem{dorney2019controlling}
K.~M. Dorney et~al.,
\newblock Nature Photonics {\bf 13}, 123 (2019).

\bibitem{schmiegelow2016transfer}
C.~T. Schmiegelow, J.~Schulz, H.~Kaufmann, T.~Ruster, U.~G. Poschinger, and
  F.~Schmidt-Kaler,
\newblock Nature Communications {\bf 7}, 12998 (2016).

\bibitem{pazourek2015attosecond}
R.~Pazourek, S.~Nagele, and J.~Burgd{\"o}rfer,
\newblock Rev. Mod. Phys. {\bf 87}, 765 (2015).

\bibitem{picon2010transferring}
A.~Pic{\'o}n, A.~Benseny, J.~Mompart, J.~R.~V. de~Aldana, L.~Plaja, G.~F.
  Calvo, and L.~Roso,
\newblock New Journal of Physics {\bf 12}, 083053 (2010).

\bibitem{picon2010photoionization}
A.~Pic{\'o}n, J.~Mompart, J.~R.~V. de~Aldana, L.~Plaja, G.~F. Calvo, and
  L.~Roso,
\newblock Optics Express {\bf 18}, 3660 (2010).

\bibitem{babiker2002orbital}
M.~Babiker, C.~R. Bennett, D.~L. Andrews, and L.~C.~D. Romero,
\newblock Physical Review Letters {\bf 89}, 143601 (2002).

\bibitem{peshkov2017photoexcitation}
A.~A. Peshkov, D.~Seipt, A.~Surzhykov, and S.~Fritzsche,
\newblock Physical Review A {\bf 96}, 023407 (2017).

\bibitem{giammanco2017influence}
F.~Giammanco, A.~Perona, P.~Marsili, F.~Conti, F.~Fidecaro, S.~Gozzini, and
  A.~Lucchesini,
\newblock Optics Letters {\bf 42}, 219 (2017).

\bibitem{watzel2016discerning}
J.~W{\"a}tzel and J.~Berakdar,
\newblock Phys. Rev. A {\bf 94}, 033414 (2016).

\bibitem{dahlstrom2013theory}
J.~M. Dahlstr{\"o}m, D.~Gu{\'e}not, K.~Kl{\"u}nder, M.~Gisselbrecht,
  J.~Mauritsson, A.~L?Huillier, A.~Maquet, and R.~Ta{\"\i}eb,
\newblock Chemical Physics {\bf 414}, 53 (2013).

\end{thebibliography}

\end{document}